%% file: main.tex
\documentclass[runningheads]{llncs}
\usepackage{graphicx}

\usepackage{tikz}
\usepackage{comment}
\usepackage{amsmath,amssymb} 
\usepackage{color}
\usepackage{xcolor}
\usepackage{times}
\usepackage{epsfig}
\usepackage{graphicx}
\usepackage{amsmath}
\usepackage{amssymb}
\usepackage{multirow}
\usepackage[T1]{fontenc}
\usepackage{mathrsfs}
\usepackage{booktabs}
\usepackage{float}

\usepackage[pagebackref=true,breaklinks=true,colorlinks,bookmarks=false]{hyperref}

\usepackage[accsupp]{axessibility}  

\begin{document}
\pagestyle{headings}
\mainmatter
\def\ECCVSubNumber{56}  

\title{Fast Nearest Convolution for Real-Time Efficient Image Super-Resolution} 

\titlerunning{Fast Nearest Convolution for Real-Time Efficient Image Super-Resolution}
%
\author{Ziwei Luo\inst{1} \and
Youwei Li\inst{1} \and 
Lei Yu\inst{1} \and
Qi Wu\inst{1} \and
Zhihong Wen\inst{1} \and \\
Haoqiang Fan\inst{1} \and  
Shuaicheng Liu\inst{2, 1}\thanks{Corresponding author.}
}
\authorrunning{Ziwei Luo et al.}
%
\institute{Megvii Technology \and
University of Electronic Science and Technology of China}

\maketitle

\input{abstract}
\input{introduction}
\input{related_work}

\input{method}

\input{experiment}

\input{conclusion}

%
%
\bibliographystyle{splncs04}
\bibliography{egbib}
\end{document}

%% file: abstract.tex
\begin{abstract}
Deep learning-based single image super-resolution (SISR) approaches have drawn much attention and achieved remarkable success on modern advanced GPUs. However, most state-of-the-art methods require a huge number of parameters, memories, and computational resources, which usually show inferior inference times when applying them to current mobile device CPUs/NPUs. In this paper, we propose a simple plain convolution network with a fast nearest convolution module (NCNet), which is NPU-friendly and can perform a reliable super-resolution in real-time. The proposed nearest convolution has the same performance as the nearest upsampling but is much faster and more suitable for Android NNAPI. Our model can be easily deployed on mobile devices with 8-bit quantization and is fully compatible with all major mobile AI accelerators. Moreover, we conduct comprehensive experiments on different tensor operations on a mobile device to illustrate the efficiency of our network architecture. Our NCNet is trained and validated on the DIV2K $3\times$ dataset, and the comparison with other efficient SR methods demonstrated that the NCNet can achieve high fidelity SR results while using fewer inference times. 
Our codes and pretrained models are publicly available at \href{https://github.com/Algolzw/NCNet}{https://github.com/Algolzw/NCNet}.

\keywords{Image super-resolution, real-time network, mobile device, nearest convolution, quantization}
\end{abstract}

%% file: introduction.tex
\section{Introduction}

Image super-resolution (SR) is a fundamental task in computer vision that aims to reconstruct high-resolution (HR) images from their degraded low-resolution (LR) counterparts. It is a hot topic in recent years since its importance and ill-posed nature. The inherent challenge in the SR problem is that there always exists infinite solutions for recovering the HR image, and different HR images can be degraded to the same LR image, which makes it difficult to directly learn the super-resolution process.

During the past decade, we have witnessed the remarkable success of deep neural network (DNN) based techniques in computer vision~\cite{krizhevsky2012imagenet,he2016deep,he2017mask,ren2015faster}. SR algorithms that are based on deep convolution networks have attracted lots of attention and rapidly developed. As a result, many works have achieved impressive results on kinds of SR tasks~\cite{timofte2017ntire,cai2019ntire,bhat2021ntire,bhat2022ntire}. However, most superior methods heavily rely on using large network capacities and model complex to improve the SR performance, which limits their practicability on real-world resource-constrained mobile devices. 

In order to apply DNN-based SR models to smartphones, a new research line called efficient super-resolution is developed where various methods have been proposed to reduce the model complexity and inference time~\cite{zhang2020aim,li2022ntire}. A representative work is the IMDN~\cite{hui2019lightweight}, which proposes an information multi-distillation block that uses feature distillation and selective fusion parts to compress the model`s parameters while preserving SR performance. Later, RFDN~\cite{liu2020residual} builds a residual feature distillation network on top of IMDN but replaces all channel splitting operations with 1$\times$1 convolutions and adds feature distillation connections. By doing so, RFDN has won 1st place in the AIM 2020 efficient super-resolution challenge~\cite{zhang2020aim}. In addition, some simplified attention mechanisms are also used in the efficient super-resolution task~\cite{zhao2020efficient,luo2020latticenet}. Compared with traditional superior SR networks, all these efficient-designed methods perform well and fast on desktop GPU devices. But we noticed that performing super-resolution on smartphones has much tighter limits on computing capacities and resources: a restricted amount of RAM, and inefficient support for many common deep learning layers and operators.

\begin{table}[t]
\centering
\resizebox{.75\linewidth}{!}
{
\begin{tabular}{lccc}
\toprule
Tensor operation node  & CPU   & GPU delegate   & Android NNAPI    \\ \midrule
Conv3 - f3-16  & 19.1ms & 13.9ms & 20.1ms   \\
w/ dilation  & 23.1ms & 27.0ms & 44.3ms   \\
+ Add  & 24.4ms & 14.7ms & 21.5ms   \\
+ Multiply  & 22.8ms & 59.8ms & 21.7ms   \\
+ Concat  & 25.5ms & - & 50.7ms   \\
+ Split  & 22.2ms & 40.4ms & 32.3ms   \\
+ ReLU  & 19.1ms & 14.5ms & 28.4ms   \\
+ LeakyReLU  & 44.7ms & 14.2ms & 66.8ms   \\
+ Global\_Avgpool  & 16.6ms & 4.9ms & 29.1ms   \\
+ Global\_Maxpool  & 102.0ms & 5.0ms & 21.1ms   \\
\bottomrule
\end{tabular}
}
\caption{Inference times of different commonly used tensor operation nodes. `w/ dilation' means a dilated convolution. `Conv\textbf{N} - f\textbf{A}-\textbf{B}' means the kernel size is \textbf{N}, input layer has \textbf{A} channels and the output layer has \textbf{C} channels.}
\label{table:tensor-op}
\end{table}

\begin{table}[t]
\centering
\resizebox{.75\linewidth}{!}
{
\begin{tabular}{lccc}
\toprule
Convolution network  & CPU   & GPU delegate   & Android NNAPI    \\ \midrule
Conv1 - f3-3  & 5.3ms & 4.7ms & 9.6ms   \\
Conv3 - f3-3  & 14.8ms & 5.0ms & 15.0ms   \\
Conv5 - f3-3  & 21.3ms & 5.8ms & 27.2ms   \\
\midrule \midrule
Conv3 - f3-8  & 14.7ms & 8.8ms & 17.4ms   \\
Conv3 - f3-16  & 19.1ms & 13.9ms & 20.1ms   \\
Conv3 - f3-32  & 27.1ms & 25.4ms & 24.7ms   \\
\midrule \midrule
Conv3 - f3-8-8  & 31.4ms & 9.8ms & 24.9ms   \\
Conv3 - f3-8-16  & 33.6ms & 15.0ms & 27.8ms   \\
Conv3 - f3-16-16  & 50.4ms & 25.8ms & 27.8ms   \\
Conv3 - f3-16-32  & 63.8ms & 32.5ms & 34.4ms   \\
Conv3 - f3-32-32  & 103.2ms & 60.7ms & 34.3ms   \\
\bottomrule
\end{tabular}
}
\caption{Inference times of different convolution network structures. `Conv\textbf{N} - f\textbf{A}-\textbf{B}-\textbf{C}' means it is a two-layer convolution network, where the kernel size is \textbf{N}, first layer has \textbf{A} channels, second layer has \textbf{B} channels, and the number of output channel is \textbf{C}.}
\label{table:conv}
\end{table}

A mobile-friendly SR model should take care of the compatibility of tensor operators on mobile NPUs. We need to know what operations are particularly optimized by the mobile NN platform (Synaptics Dolphin platform). And the same tensor operation could have different inference times on different smartphone AI accelerators (e.g., CPU, GPU delegate, and Android NNAPI). Recent works~\cite{du2021anchor,ayazoglu2021extremely,ignatov2021real} have investigated some limiting factors of running deep networks on a mobile device and what kind of architecture can be friendly to INT8 quantization. They propose several useful techniques such as ``anchor-based residual learning'', ``Clipped ReLU'', and ``Quantize-Aware Training'' to accelerate inference while preserving accuracy on smartphones. However, there is still no basic experiment that can illustrate the difference of tensor operators on different smartphone AI accelerators.

In this paper, we provide a comprehensive comparison of inference times for kinds of tensor operation nodes and network architectures, as shown in Table~\ref{table:tensor-op} and Table~\ref{table:conv}. We use DIV2K 3$\times$~\cite{timofte2017ntire} as the training and evaluation dataset. All experiments are evaluated on \textit{AI Benchmark} application~\cite{ignatov2018ai,ignatov2019ai}. As one can see, some commonly used deep learning techniques (e.g., dilated convolution, concatenation, channel splitting, and LeakyReLU) are not compatible with mobile Android NNAPI, even though they have a good performance on CPU and GPU delegate. Based on these experiments and analysis, we design a plain network that only contains 3$\times$3 convolution layers and ReLU activation functions. Moreover, we propose to use a novel nearest convolution to replace the traditional nearest upsampling in network residual learning, which further speeds up the inference and achieves the same effect as nearest interpolation residual learning. In summary, our main contributions are as follows:

\begin{itemize}
    \item We provide a comprehensive comparison of inference times for different tensor operators and network architectures on a smartphone, which tells us what operation is good for mobile devices and should be incorporated into the network.
    \item We propose a fast nearest convolution plain network (NCNet) that is mobile-friendly and can achieve the same performance as nearest interpolation residual learning while saving approximately 40ms on a Google Pixel 4 smartphone.
\end{itemize}

%% file: related_work.tex
\section{Related Work}

\subsection{Single Image Super-Resolution}

Single Image Super-Resolution (SISR) is one of the most popular research topics in computer vision due to its importance and ill-posed nature. The pioneering deep learning based method is SRCNN~\cite{simonyan2014very} which applies the bicubic downsampling on HR images to construct HR and LR pairs and employs a simple 3-layers convolution neural network (CNN) to perform super-resolution. After then, various super-resolution approaches are proposed and have achieved remarkable performance in processing the SISR problem~\cite{kim2016accurate,lim2017enhanced,shi2016real,zhang2018residual,haris2018deep,tai2017image,luo2021ebsr,zhang2020residual,luo2022deep,luo2022bsrt}. For example, VDSR~\cite{simonyan2014very} proposes a very deep CNN to improve the SR results and ESPCN~\cite{shi2016real} designs a simple yet efficient strategy, called pixel-shuffle, for real-time feature upsampling. EDSR~\cite{lim2017enhanced} proposes to enhance the CNN-based SR network by removing the batch normalization layer of all residual blocks. Moreover, to improve the perceptual visual quality, recent works~\cite{ledig2017photo,sajjadi2017enhancenet,wang2018esrgan} propose to employ some advanced losses such as the VGG loss~\cite{simonyan2014very}, perceptual loss~\cite{johnson2016perceptual}, and GAN loss~\cite{goodfellow2014generative} to help the network to learn realistic image details. However, these methods usually require huge memories and computational resources which makes them hardly be applied to modern mobile devices.

\subsection{Efficient Image Super-Resolution}

To fit the growing demands of deploying models on real-world smartphone applications, many works have refocused their attention on efficient image super-resolution techniques~\cite{ahn2018fast,dong2016accelerating,hui2019lightweight,liu2020residual,tai2017image}. CARN~\cite{ahn2018fast} uses cascaded residual blocks and group convolution to achieve a lightweight SR network. IMDN~\cite{hui2019lightweight} designs an information multi-distillation network that extracts hierarchical features and expresses the number of filters in each block to reduce the memories and FLOPs. The following work RFDN~\cite{liu2020residual} further improves the network by introducing feature distillation blocks that employ 1$\times$1 convolution layers to replace all channel splitting operations. Although these networks can perform efficient SR on desktop CPUs/GPUs, they are still not feasible to be deployed on real-world applications since the computational resources on most smartphones are much lower than on computers. Address it, ABPN~\cite{du2021anchor} and XLSR~\cite{ayazoglu2021extremely} are winners of the Mobile 2021 Real-Time Single Image Super-resolution Challenge~\cite{ignatov2021real}. They investigated some limiting factors of the mobile device models and proposed extremely lightweight SR networks for the mobile SR problem. Moreover, the INT8 quantization is widely used in mobile devices since it can accelerate inference and save memories, as illustrated in Table \ref{table:quantization} (for runtime we mainly focus on Android NNAPI).

\begin{table}[bh]
\centering
\resizebox{.76\linewidth}{!}
{
\begin{tabular}{cccccc}
\toprule
Quantization & \#Params & CPU   & GPU delegate   & NNAPI   & PSNR-int8   \\ \midrule
float-32 &43K & 506ms & - & 153ms & 30.21   \\
int-8 &43K & 509ms & - & 135ms & 30.06   \\
\bottomrule
\end{tabular}
}
\caption{Runtime and PSNR comparison on different quantization modes of ABPN~\cite{du2021anchor}.}
\label{table:quantization}
\end{table}

%% file: method.tex
\begin{figure}[t]
\begin{center}
\includegraphics[width=1.\linewidth]{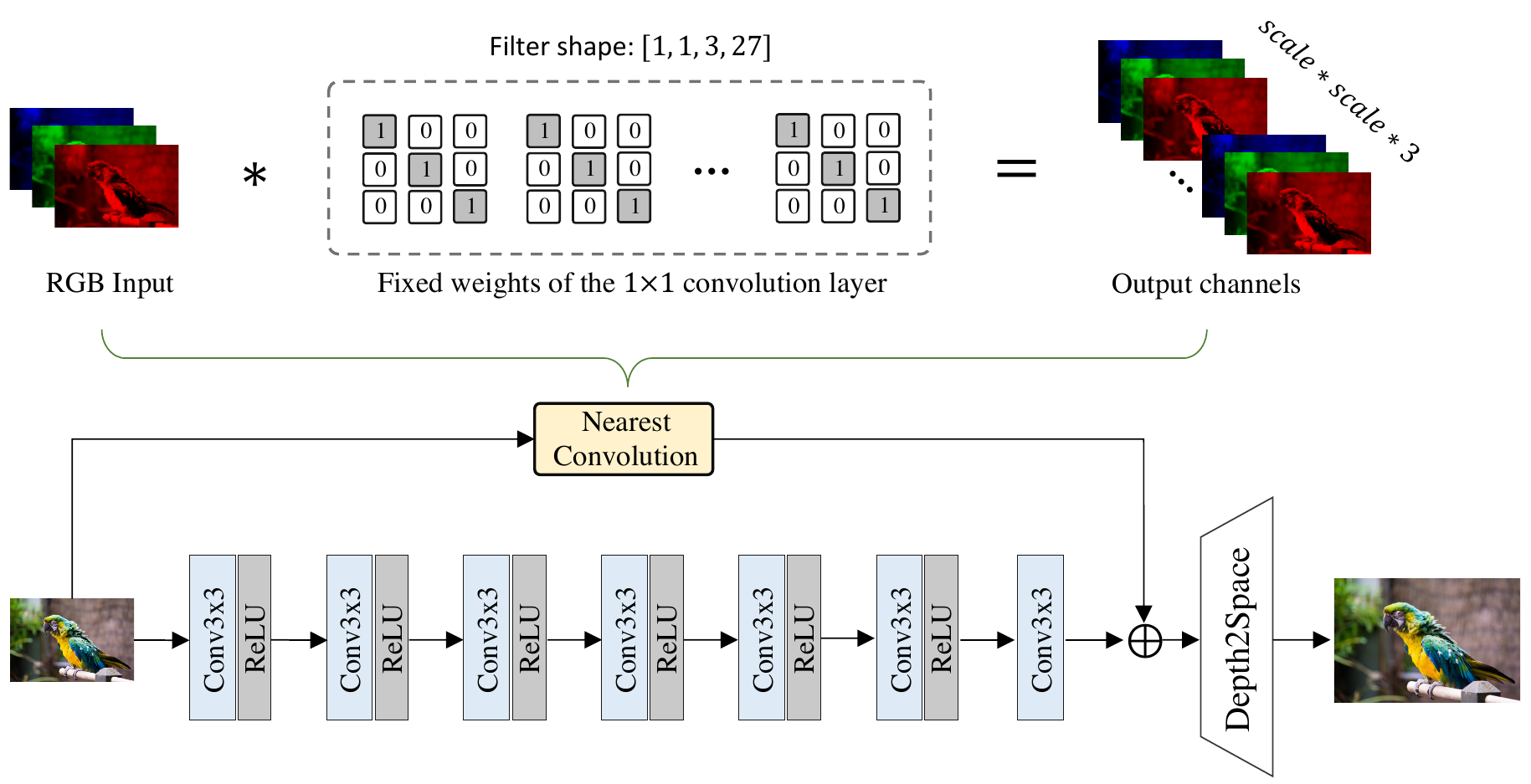}
\end{center}
    \caption{Network architecture of the proposed NCNet. The main network backbone learns the residual while the nearest convolution module directly delivers the low-frequency information to the final result. Moreover, the nearest convolution achieves the same performance as nearest interpolation but can be executed parallelly.}
\label{fig:overview}
\end{figure}

\section{Method}

In this section, we will start by finding a proper network architecture for the mobile device (especially for the Android NNAPI accelerator), based on Table~\ref{table:tensor-op} and Table~\ref{table:conv}. Then we describe the main idea of the nearest convolution module. By assembling the manually designed backbone and the nearest convolution in the form of residual learning, we can obtain the final efficient SR network, as shown in Figure~\ref{fig:overview}.

\subsection{Network Architecture Selection}

To build a real-time mobile-friendly SR network, our first thing is to figure out what tensor operators are compatible and efficient for the Android NNAPI accelerator. To make use of the NPU's parallel property, the baseline operation node is set to a 3$\times$3 convolution with 16 output channels, so we could add kinds of multi-channel (element-wise) tensor operations to it as in Table~\ref{table:tensor-op}. In addition, we also want to know which convolution architecture and channel are better. So we compared the arrangement and combination of convolution layers in Table \ref{table:conv}. Note that all experiments are based on a Google Pixel 4 smartphone, using INT8 quantization. 

From our observation, we find that the inference times of channel splitting and concatenation on NPUs are much lower than on CPUs, which means these operators are not friendly to NPU parallelism, and the runtime will exponentially increase if the number of channels doubles. This situation also happens on `LeakyReLU' and `Global Average Pooling'. In recent years, many advanced methods like using LeakyReLU as their default activation function~\cite{wang2018esrgan}, but it is obviously not suitable for mobile NPUs. Similarly, the attention mechanism is widely used in state-of-the-art SR approaches~\cite{dai2019second,zhang2018image} but it is time-consuming and not a good choice for mobile devices. For the choice of convolution layers, we find the 5$\times$5 convolution is wasteful and inefficient, and the 1$\times$1 convolution is not convincing to achieve a good performance. Thus we choose to set the kernel size to 3$\times$3 for all convolution layers. For multi-layer structures, we surprisingly find that keeping all layers the same number of channels has approximately the same runtime as changing channels. Note the latter has fewer parameters, which means we could use a larger network to achieve better performance while preserving the same inference time on mobile NPUs. Therefore, the main backbone of our network is designed to only contain 3$\times$3 convolution layers with ReLU activation functions. And inspired by ABPN~\cite{du2021anchor}, our solution also incorporates the residual learning of RGB images to improve the final result. The overview of the proposed fast Nearest Convolution (NCNet) network is shown in Figure \ref{fig:overview}.

\subsection{Nearest Convolution}

The most important component of the NCNet is the Nearest Convolution module, which is actually a special $1 \times 1$ convolution layer with stride 1. To achieve the nearest interpolation, the weights of the convolution are freezed and manually filled by $s^2$ groups of $3 \times 3$ identity matrix (where $s$ is the upscale factor) and each group would produce an RGB image, which just like a copy operation to repeat the input image $s^2$ times. Then these $s^2$ RGB images can reconstruct an HR image through a depth-to-space operation. In this way, the reconstructed image will be exactly the same as the nearest interpolated HR image but is much faster especially using mobile GPUs/NPUs. The reason is that when the 1$\times$1 nearest convolution is performed on the NPU devices, it can be executed in parallel thus showing superior to other normal interpolation operations. The inference time of different upsampling methods is shown in Table \ref{table:compare_upsample}. As one can see, the proposed nearest convolution can save approximately 40ms compared with the original nearest upsampling operation. 

\subsection{Residual Learning}
In practice, we'd like to add these $s^2$ RGB images to the output of the plain network before the depth-to-space layer then the plain network could focus on learning the residual information. Let ${\bf x}$ be the HR image and ${\bf y}$ be its degraded LR image. We could obtain the super-resolved image $\hat{\bf x}$ by:
\begin{equation}
\centering
    \hat{\bf x} = f({\bf y};\theta) = D2S(f_{res}(y;\theta) + f_{nc}(y)),
\label{eq:reslearn}
\end{equation}
where $f(\cdot)$ represents the SR network and $\theta$ is the network's parameters. $f_{res}(\cdot)$ and $f_{nc}(\cdot)$ represent the residual learning network and nearest convolution, respectively. $D2S$ means the depth-to-space layer.
To illustrate the effectiveness of our network, we use L1 loss to optimize our model, which is formulated as follows:
\begin{equation}
\centering
    {\cal L}_1(\theta) =  \frac{1}{N}\sum_{i=1}^{N}(f({\bf y};\theta) - {\bf x}).
\label{eq:loss}
\end{equation}
To accelerate the inference time, we only incorporate $7$ layers of $3 \times 3$ convolution with the ReLU activation function for the whole network, and the number of channels is fixed to $32$. Since the runtime of element-wise operation is non-negligible, we didn't use the residual connection for each convolution layer.

%% file: experiment.tex
\begin{table}[t]
\centering
\resizebox{.85\linewidth}{!}
{
\begin{tabular}{ccccc}
\toprule
Upsample methods & CPU   & GPU delegate   & NNAPI    & PSNR  \\ \midrule
nearest & 23.1ms & \textbf{19.0ms} & 55.0ms & 26.67  \\
bilinear & 77.7ms & 21.0ms & 128.2ms & \textbf{27.67}  \\
Conv-3 + depth2space & 30.8ms & 26.5ms & 43.8ms & - \\
nearest convolution + depth2space & \textbf{15.9ms} & 20.3ms & \textbf{14.8ms} & 26.67 \\
\bottomrule
\end{tabular}
}
\caption{The table shows the inference time between the proposed nearest convolution and other commonly used upsample methods on different mobile accelerators. The proposed nearest convolution can save approximately 40ms compared with the original nearest upsampling operation.}
\label{table:compare_upsample}
\end{table}

\begin{table}[t]
\centering
\resizebox{.99\linewidth}{!}
{
\begin{tabular}{cccc|ccc}
\toprule
Google Pixel 4 & CPU   & GPU delegate   & NNAPI   & \#Params   & PSNR-float32 & PSNR-int8  \\ \midrule
NCNet & 535.1ms & 263.0ms & 104.0ms & 53K & 30.27dB & 30.18dB  \\
\bottomrule
\end{tabular}
}
\caption{The table shows the runtime on different accelerators and the PSNR performance of the proposed NCNet on a Google Pixel 4 smartphone.}
\label{table:compare_time}
\end{table}

\begin{table}[t]
\centering
\resizebox{.8\linewidth}{!}
{
\begin{tabular}{cccccc}
\toprule
Method & \#Params & CPU   & GPU delegate   & NNAPI   & PSNR-int8   \\ \midrule
FSRCNN~\cite{dong2016accelerating} &24K & 476ms & 226ms & 251ms & 28.34   \\
ABPN~\cite{du2021anchor} &43K & 509ms & - & 135ms & 30.15   \\
NCNet(ours)  & 53K  & 535ms & 263ms & 104ms & 30.18   \\
\bottomrule
\end{tabular}
}
\caption{Runtime and PSNR comparison with FSRCNN~\cite{dong2016accelerating} and ABPN~\cite{du2021anchor}. Note the ABPN fails to run on the mobile GPU delegate.}
\label{table:compare_other}
\end{table}

\section{Experiments}
\subsection{Dataset and Implementation Details}
We use DIV2K~\cite{agustsson2017ntire} as the training (800 image pairs) and testing (100 image pairs)  dataset. The scale factor is fixed to $3$ and the batch size is set to 64. The patch size of LR images is 64 and the total iterations are set to 500,000. All parameters are initialized using Xavier initializer~\cite{glorot2010understanding}. We use the Adam optimizer~\cite{kingma2014adam}, where the initial learning rate is $1\times 10^{-3}$ and decreases by half every 200,000 iterations. Inspired from~\cite{li2022ntire}, we also finetune the trained model with a larger LR patch size of 128 for additional 200,000 iterations. In this paper, we follow the Mobile AI \& AIM 2022 Real-Time Image Super-Resolution Challenge~\cite{ignatov2022isr} to measure super-resolved results in the RGB space. Moreover, we use a single NVIDIA RTX 2080Ti with 8 CPUs to train and evaluate the original non-quantized model.

\subsection{Model Quantization}
For network quantization, we use the standard Tensorflow Quantization tool - TFLite - to quantize the trained model as the Post-Quantization strategy. To evaluate the PSNR of the quantized models (float32 and int8), we set the input shape to $[1, None, None, 3]$ to allow super-resolving arbitrary shape images. To evaluate the inference time, we fix the input shape to $[1, 360, 640, 3]$ and test the model in the \textit{AI Benchmark}~\cite{ignatov2018ai} application on a Google Pixel 4 smartphone. 

\begin{figure}[H]
\begin{center}
\includegraphics[width=1.\linewidth]{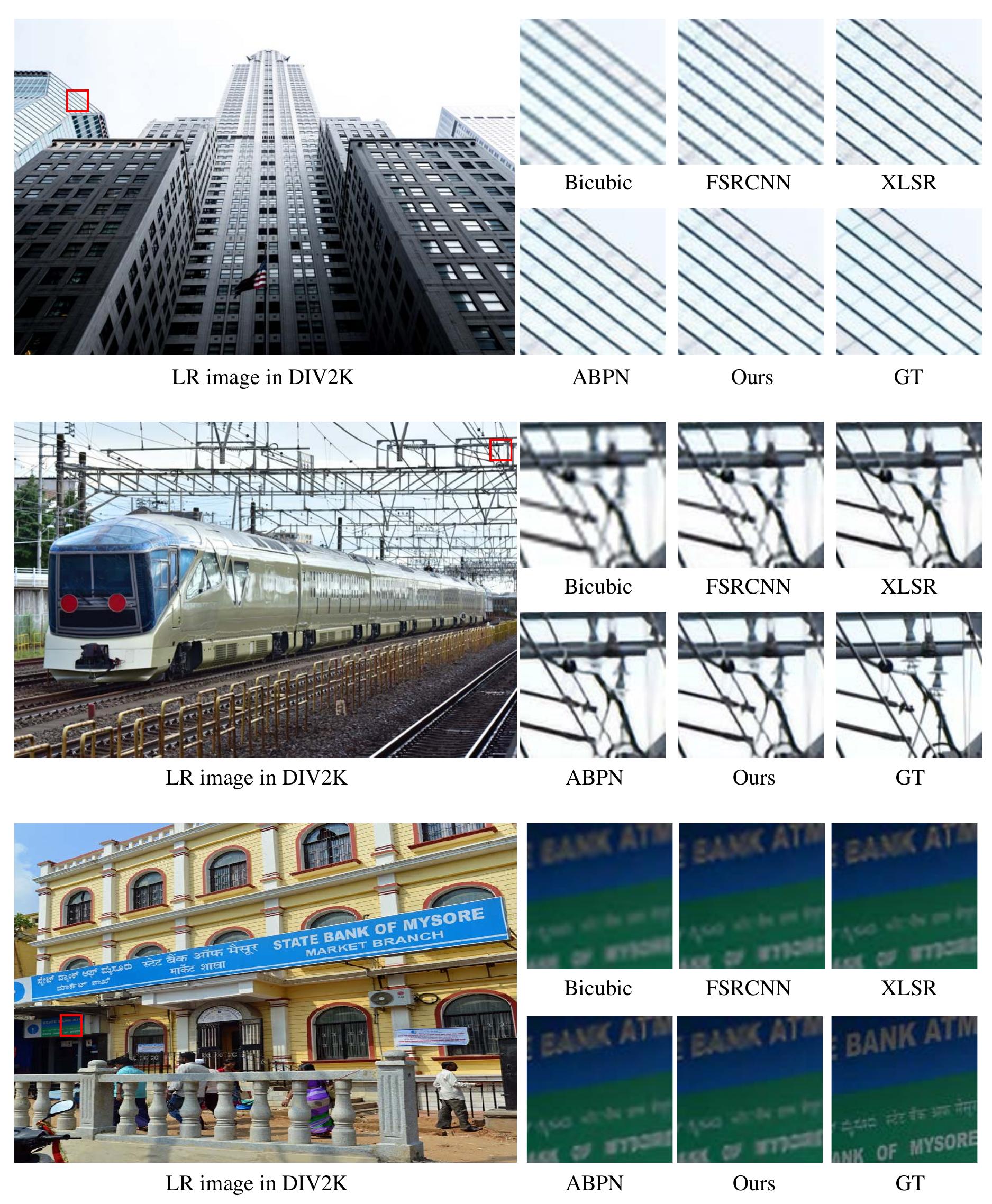}
\end{center}
    \caption{Visual comparison of LR Img 846, Img 820 and Img 819 in DIV2K validation dataset~\cite{agustsson2017ntire}. All results are produced by the INT8 quantization model, for scale factor 3.}
\label{fig:compare}
\end{figure}

\subsection{Experimental Results}
After training and quantizing, we can show the overall information of the proposed model in Table~\ref{table:compare_time}. As one can see, our NCNet is NPU-friendly. The runtime on NNAPI is 5$\times$ faster than CPU and 2.5$\times$ than GPU. With 53K parameters, the INT8 quantized NCNet only loses 0.09dB on the mobile device compared with its float32 model.

\begin{figure}[ht]
\centering
\includegraphics[width=.6\linewidth]{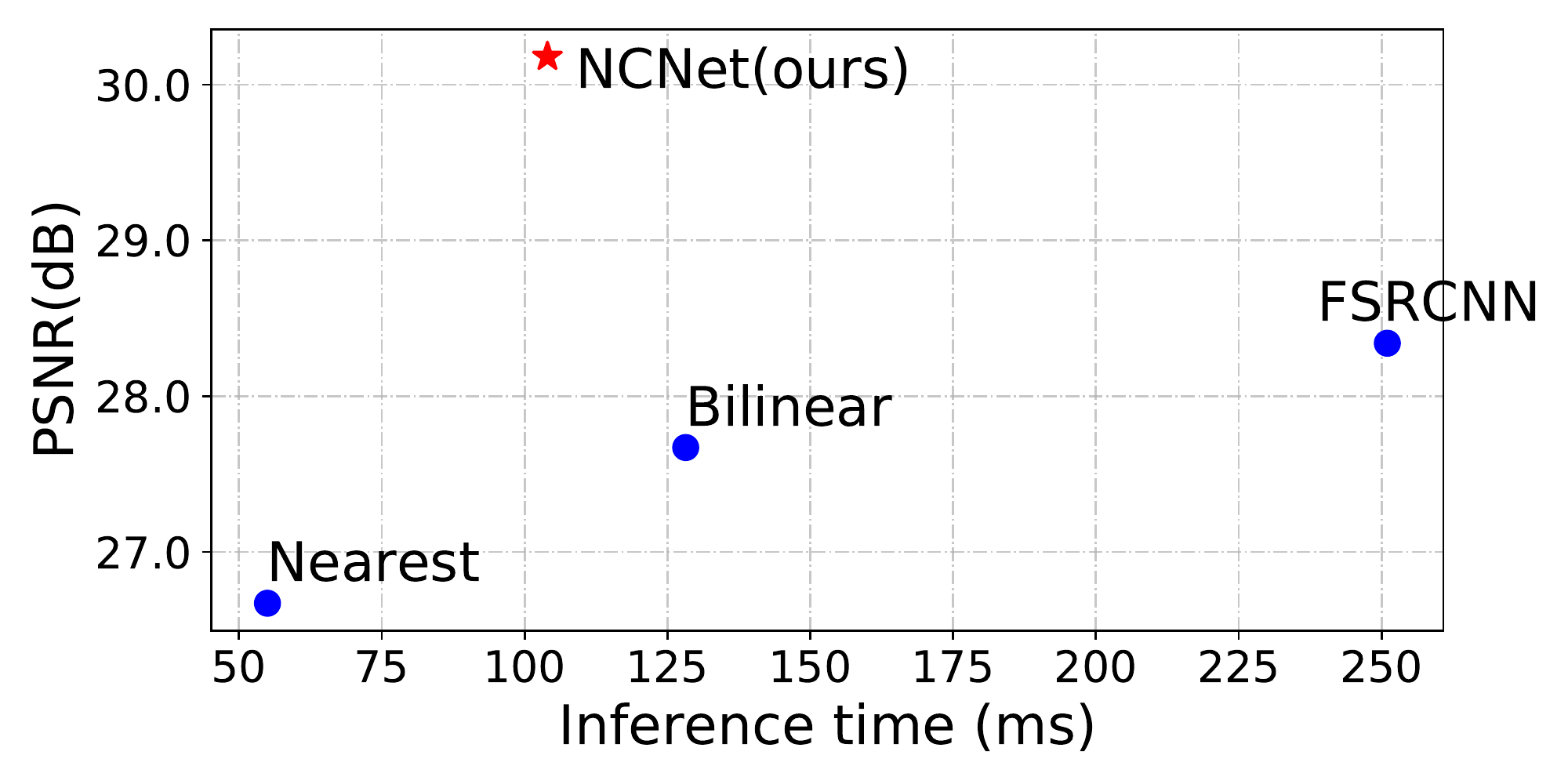}
\caption{Comparison of PSNR and inference time on Google Pixel 4 with INT8 quantization.}
\label{fig:psnr}
\end{figure}

We compare the proposed NCNet with other lightweight real-time SR algorithms, such as FSRCNN~\cite{dong2016accelerating} and ABPN~\cite{du2021anchor}. The former is the pioneering DL-based SR network and the latter is a superior method in Mobile AI 2021 Real-Time Single Image Super Resolution Challenge~\cite{ignatov2021real}. The quantitative results are illustrated in Table~\ref{table:compare_other}. The proposed NCNet is faster than ABPN and FSRCNN on Android NNAPI and also achieves the best PSNR performance. In addition, we also compared two classical upsampling methods: Nearest upsampling and Bilinear upsampling. Both of them can be quantized to INT8 and are well compatible with Android NNAPI. The result is illustrated in Figure~\ref{fig:psnr}. Our method is even faster than bilinear upsampling while achieving an impressive performance.

The visual comparison of our method and other approaches with INT8 quantization is shown in Figure~\ref{fig:compare}. The proposed NCNet has more textures and can produce visually pleasant SR images.


%% file: conclusion.tex
\section{Conclusion}
In this paper, we introduce an efficient fast nearest convolution network (NCNet) for real-time super-resolution. It is well compatible with INT8 quantization and Android NNAPI accelerator. By assembling the CNN-based plain network and the nearest convolution as residual learning architecture, NCNet achieves a remarkable performance while preserving real-time inference. By utilizing the NPU's parallel property, our model's runtime on the Android NNAPI is even faster than the traditional bilinear upsampling. We also provide a comprehensive comparison of inference times for different tensor operators and network architectures on the smartphone, which could help to select operators and architectures for real-world mobile devices.